\documentclass[a4paper]{article}
\usepackage[T2A]{fontenc}
\usepackage{amsfonts,amssymb,amsmath,amsthm,cite}
\usepackage{ucs}
\usepackage[utf8x]{inputenc}
\usepackage{graphicx}

\usepackage{slashed}
\usepackage{textcomp}
\usepackage[notref,notcite]{}

\evensidemargin=\oddsidemargin
\newtheorem{assumption}{Assumption}[section]

\newtheorem{thm}[assumption]{Theorem}

\newtheorem{lemma}[assumption]{Lemma}

\begin{document}
\begin{center}
	
	{\Large\textbf{Mellin--Barnes transformation for two-loop master-diagram}}
	\vspace{0.5cm}
	
	{\large S.~E.~Derkachev$^\ddagger$\,,\,\,\,\,A.~V.~Ivanov$^\dag$\,,\,\,\,\,L.~A.~Shumilov$^\sharp$}
	
	\vspace{0.5cm}
	
	$^\ddagger$$^\dag${\it St. Petersburg Department of the Steklov Mathematical Institute of Russian Academy of Sciences,}\\{\it Fontanka 27, 191023 St. Petersburg, Russia}\\
	$^\sharp${\it St. Petersburg Academic University,}\\{\it Khlopina str. 8/3, St. Petersburg, 194021, Russia}\\
\end{center}
	$^\ddagger${\it E-mail: derkach@pdmi.ras.ru}\\
	$^\dag${\it E-mail: regul1@mail.ru}\\
	$^\sharp${\it E-mail: la\_shum@mail.ru}	

\renewcommand{\abstractname}{Abstract} 
\renewcommand{\figurename}{Fig.}
\renewcommand{\refname}{Reference list}

\begin{abstract}
In the paper, we obtain an expression for a two-loop master-diagram by using the Mellin–-Barnes transformation. In the two-dimensional case we managed to factorize the answer and write it as a bilinear combination of hypergeometric functions ${}_3F_2$.
\end{abstract}

\section{Introduction}

The method of separation of variables, developed by E.K. Sklyanin~\cite{S1,S2,S3}, is one of the effective methods in the theory of quantum integrable models (see reviews \cite{F,KS,S1}).
It is based on the transition from the coordinate representation to the separated variables representation, in which the problem of studying the original Hamiltonian is drastically simplified.
Such a unitary transformation is carried out by an integral operator, the kernel of which is the mutual eigenfunctions of a special family of commuting operators. The quantum Toda chain~\cite{S2,KL,GKL} and the two-dimensional model of the non-compact spin chain~\cite{DKM} are two meaningful examples of quantum integrable models where the program of separation of variables was fully implemented. 

In the works of L.N. Lipatov~\cite{L1,L2}, L.D. Faddeev, and G.P. Korchemskiy~\cite{FK}, it was discovered that the two-dimensional model of a non-compact spin chain describes high-energy processes in quantum chromodynamics. The operator of transition to the representation of separated variables for this model was constructed explicitly in~\cite{DKM}.
In the simplest case, either a two-dimensional Fourier transform or a two-dimensional generalization of the Mellin transform is obtained, so the general case is a far-reaching generalization of the two classical integral transforms.
The transition to separated variables turned out to be a useful tool for calculating multiloop Feynman diagrams in quantum field theory. Basso and Dixon's work \textit{"Gluing Ladder Feynman Diagrams into Fishnets"}~\cite{BD} proposed a remarkable determinant representation of a special class of Feynman diagrams, which play a dominant role in the four-dimensional Fishnet CFT model~\cite{GK} as a hypothesis. Then, in~\cite{DKO}, using an integral transformation to separated variables, an analogue of determinant Basso--Dixon representation was derived for the diagrams in two-dimensional Fishnet CFT~\cite{KO}, and in~\cite{DO} a four-dimensional generalization was obtained.
Thus, the Basso--Dixon conjecture was proved, and the main tool used was the transformation to separated variables. The necessary generalization of the basic formulae~\cite {DKM} from two-dimensional to four-dimensional case was developed in~\cite {DO}. 

An important open question is the generalization to an arbitrary number of dimensions, but in the simplest case the necessary formulae are quite obvious and are known~\cite {KO, BFKZ}. By analogy with the two-dimensional case, we will call the simplest integral transform the Mellin--Barnes one. In this work the Mellin--Barnes transform is used to calculate the two-loop master diagram~\cite {Gr, BW}. 

The article is organized as follows. In the second section a general calculation scheme is formulated. In the third section the scheme is applied to the calculation of the two-loop diagram in the two-dimensional case. The last section is devoted to the case of an arbitrary number of dimensions. In the Appendix we give a proof of the unitarity of the Mellin--Barnes transform for an arbitrary number of dimensions. 

The two-dimensional case is highlighted for two reasons. First, in the case, all calculations can be carried out for a diagram with arbitrary tensor lines, not being limited only to scalar propagators. Secondly a remarkable factorization occurs: a two-fold sum is factored into the product of two one-fold sums.

\section{Formularion of the problem}

In this section we will formulate the problem schematically. Suppose $\mathcal{N}$ is an abstract set of indices. Elements of the set $\alpha\in\mathcal{N}$ can have a complex structure, including continuous and discrete parameters. Let $\mathcal{V}$ be the region of integration and let the variables $z$ be functions $\mathcal{V}\to\mathbb{C}$. We will notate the integration measure as $\mathcal{D}z$.
Let us consider the family of functions
\begin{equation}\label{family}
D^{\alpha}(z_1,z_2):\,\mathcal{V}\times\mathcal{V}\to\mathbb{\overline{C}},\,\,\,\alpha\in\mathcal{N}.
\end{equation}

\begin{figure}[h]
	\centerline{\includegraphics[width=0.32\linewidth]{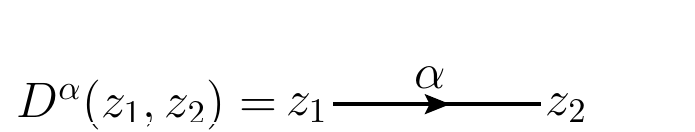}}
	\caption{The <<line>>}
	\label{line}
\end{figure}

Such functions are the basic elements of diagram technique and is denoted by a directed line from point $z_1$ to point $z_2$ with index $\alpha$, as shown in Fig. \ref{line}. We denote the integration over $\mathcal{V}$ with measure $\mathcal{D}z$ by a point.

\begin{figure}[h]
	\centerline{\includegraphics[width=0.25\linewidth]{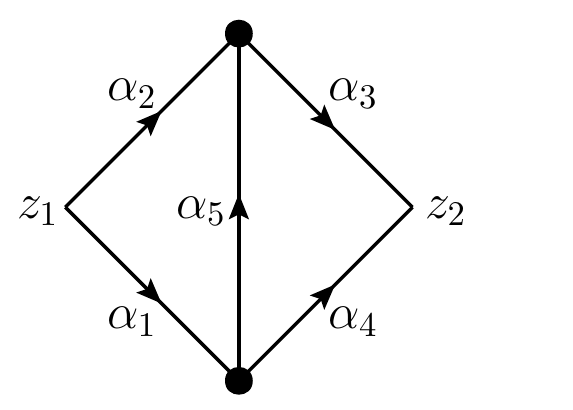}}
	\caption{Two-loop diagram.}
	\label{dia}
\end{figure}

The subject of this work is shown in Fig. \ref{dia}. Let us note that such integral
 is understood in the sense of analytic continuation from the chosen region of convergence to the complex plane as a function of external parameters $\alpha_1,\ldots,\alpha_5$.

\begin{figure}[h]
	\centerline{\includegraphics[width=0.5\linewidth]{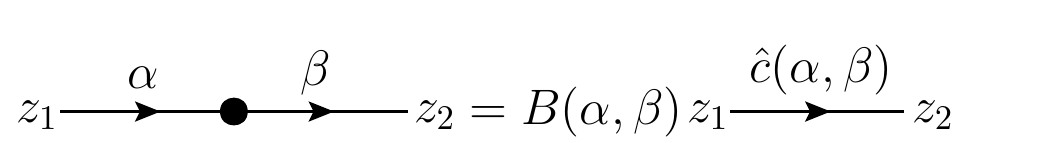}}
	\caption{The chain rule.}
	\label{chain}
\end{figure}

\begin{figure}[h]
	\centerline{\includegraphics[width=0.5\linewidth]{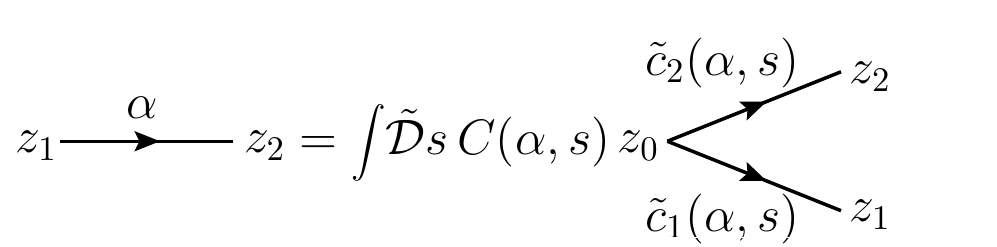}}
	\caption{Mellin--Barnes representation.}
	\label{mb}
\end{figure}

As a rule, in the theory of integrable  models the functions $D^{\alpha}$ have additional properties that simplify the problem and let us write out an explicit answer. We highlight the three main ones:
\begin{enumerate}
	\item Let $\alpha,\beta\in\mathcal{N}$, then
	\begin{equation}
	\label{mult}
	D^{\alpha}(z_1,z_2)D^{\beta}(z_1,z_2)=D^{c}(z_1,z_2),
	\end{equation}
	where $c\in\mathcal{N}$ is a function of parameters $\alpha$ and $\beta$;
	\item Let $\alpha,\beta\in\mathcal{N}$, then the chain rule is shown in Fig. \ref{chain}, where $\hat{c}\in\mathcal{N}$ is a function of $\alpha$ and $\beta$;
	\item Let $\alpha \in \mathcal{N}$, then the Mellin--Barnes representation is shown in Fig. \ref{mb}, where $\mathcal{\tilde{D}}s$ is measure on the set $\mathcal{N}$, $z_0$ is an arbitrary parameter, $\tilde{c}_1\in\mathcal{N}$ and $\tilde{c}_2\in\mathcal{N}$ are functions of the parameters $\alpha$ and $s$.
\end{enumerate}

In what follows we assume that all of these properties are satisfied. They allow us to easily get the answer for the diagram. Indeed, applying Mellin--Barnes transform to the diagonal with index $\alpha_5$ shown in Fig. \ref{dia}, choosing arbitrary parameter $z_0$ equals to $z_1$ (or $z_2$) and applying property (\ref{mult}) twice and then the chain rule, we get the answer in the following form
\begin{equation}
\label{int}
D^{b}(z_1,z_2) = \int\tilde{\mathcal{D}}s\,C(\alpha_5,s)B(b_1(s),\alpha_4)B(b_2(s),\alpha_3),
\end{equation}
where $b_1(s)=c(\alpha_1,\tilde{c}_1(\alpha_5,s))$, $b_2(s)=c(\alpha_2,\tilde{c}_2(\alpha_5,s))$, and $b=c(\hat{c}(b_1,\alpha_4),\hat{c}(b_2,\alpha_3))$.

In the rest of the work we present two special cases (two-dimensional and multidimensional). In the first case properties 1) –3) are already known and the main result is the calculation of integral (\ref{int}) and further factorization of the answer. The second case is devoted to the study of the above properties 2) and 3), as well as the calculation of the resulting integral by residues.

\section{Two-dimensional case}

In this case the set of indices $\mathcal{N}$ is  $\mathbb{C}$. Moreover, it is convenient to define <<conjugate>> index $\overline{\alpha}$ for each  $\alpha\in\mathbb{C}$ by the rule $\alpha-\overline{\alpha}=n_{\alpha}\in\mathbb{Z}$.
For convenience, let us also define auxiliary functions 
\begin{equation}
\label{function}
a(\alpha)=\frac{\Gamma(1-\overline{\alpha})}{\Gamma(\alpha)}
\,\,\,\,\,\,\mbox{and}\,\,\,\,\,\,
a(\overline{\alpha})=\frac{\Gamma(1-\alpha)}{\Gamma(\overline{\alpha})},
\end{equation}
where $\Gamma$ is Euler's Gamma function. For brevity the product of such objects we denote by $a(\alpha_1,\ldots,\alpha_n)=a(\alpha_1)\cdots a(\alpha_n)$. One can verify that such coefficients have the properties
\begin{align*}
a(\alpha)a(1-\overline{\alpha})=1,\,\,\,\,\,\,
a(\alpha)a(1-\alpha)=(-1)^{n_\alpha},
\end{align*}
and have simple poles, distributed according to the following lemma.
\begin{lemma}
	\label{l1}
	Let $n\in\mathbb{N}$, $m\in\mathbb{Z}$, $x,y\in\mathbb{R}$, $\nu\in\mathbb{C}$,
	$s=\frac{n}{2}+i\nu$,
	$\overline{s}=-\frac{n}{2}+i\nu$, $\overline{\alpha}=x+iy$, and
	$\alpha=x+m+iy$, then function $a(1-\alpha\pm s)$ has simple poles with respect to the variable $\nu$ at the points
	\begin{equation*}
	\nu=\mp i\left(\overline{\alpha}+\frac{m+\mid n\mp m\mid}{2}+k\right),\,k\in\mathbb{N}
	\cup\{0\}.
	\end{equation*}
\end{lemma}
\noindent\textbf{Proof:}
It is sufficient to consider the case with the plus sign. Let us introduce auxiliary variables
\begin{equation*}
n=n'+m,\,\nu=ip+y,\,b=x+\frac{m}{2},
\end{equation*}
then we can rewrite the function in the following form:
\begin{equation*}
a(1-\alpha+s)=\frac{\Gamma(\overline{\alpha}-\overline{s})}
{\Gamma(1-\alpha+s)}=
\frac{\Gamma(x+iy+\frac{n}{2}-i\nu)}
{\Gamma(1-m-x-iy+\frac{n}{2}+i\nu)}=
\frac{\Gamma(b+\frac{n'}{2}+p)}
{\Gamma(1-b+\frac{n'}{2}-p)}.
\end{equation*}
Taking into account the poles of the upper Gamma function and the zeros of the lower one,
we can conclude that singularities exist at the points
$p=-b-\frac{\mid n'\mid}{2}-k$,
$k\in\mathbb{N}\cup\{0\}$ with respect to the variable  $p$. The transition to the original variables proves the lemma. $\blacksquare$\\

The integration region $\mathcal{V}$ coincides with $\mathbb{C}$, the
variables are complex numbers, and the integration measure has the form $\mathcal{D}z=d\,\mathrm{Re}(z)d\,\mathrm{Im}(z)$.

Let $\nu\in\mathbb{R}$, $n\in\mathbb{Z}$,
$s=\frac{n}{2}+i\nu$, and $\overline{s}=-\frac{n}{2}+i\nu$, then, taking into account the definitions
\begin{equation*}
\int\limits \tilde{\mathcal{D}}s=
\sum\limits_{n\in\mathbb{Z}}\int_{\mathbb{R}}d\nu\,\,\,\,\,\,\mbox{and}\,\,\,\,\,\,
D^{\alpha}(z_1,z_2)=(z_2-z_1)^{\alpha}(\overline{z}_2-\overline{z}_1)^{\overline{\alpha}},
\end{equation*}
the remaining coefficients and functions are written as:
\begin{equation}
c(\alpha,\beta)=\alpha+\beta,
\end{equation}
\begin{equation}
B(\alpha,\beta)=(-1)^{n_{\alpha}+n_{\beta}}\pi a(\alpha,\beta,2-\alpha-\beta),\,\,
\hat{c}(\alpha,\beta)=\alpha+\beta-1,
\end{equation}
\begin{equation}
C(\alpha,s)=\frac{(-1)^{n_{\alpha}+n_{s}}}{2\pi}a(\alpha,1-s,1+s-\alpha),\,\,
\tilde{c}_1(\alpha,s)=s,\,\,
\tilde{c}_2(\alpha,s)=\alpha-s.
\end{equation}

Then formula (\ref{int}) has the form
\begin{multline}
\label{int1}
\frac{\pi}{2}D^{\alpha_{12345}-2}(z_1,z_2)
a(\alpha_3,\alpha_4,\alpha_5)(-1)^{n_{\alpha_{1234}}}\\
\int_{\gamma}\tilde{\mathcal{D}}s\,
a(1-s,\alpha_{25}-s,2-\alpha_{14}-s)\,
a(1+s-\alpha_5,\alpha_1+s,2-\alpha_{235}+s)(-1)^n,
\end{multline}
where the contour $\gamma$ coincides with $\mathbb{R}$ except for the point $\nu=0$, where it goes around from below.
For simplicity we use the following short notation for the sums
$\alpha_{ik} = \alpha_i +\alpha_k\,,
n_{\alpha_{ik}} = n_{\alpha_{i}} + n_{\alpha_{k}}$,
$\alpha_{ikj} = \alpha_i +\alpha_k+\alpha_j\,,
n_{\alpha_{ikj}} = n_{\alpha_{i}} + n_{\alpha_{k}}+ n_{\alpha_{j}}$ etc.

Obviously, the integrand has six series of poles in the complex plane, three of which (with the argument~$−s$) go to infinity in the upper half-plane, and the rest -- in the lower one.

For convenience, we introduce a set of auxiliary notations:
\begin{equation*}
\beta_2=1-\alpha_{25},\,\,\,\,\,\,
\beta_3=-1+\alpha_{14},
\end{equation*}
\begin{equation*}
\beta_4=\alpha_5,\,\,\,\,\,\,
\beta_5=1-\alpha_1,\,\,\,\,\,\,
\beta_6=-1+\alpha_{235},
\end{equation*}
then, taking into account Lemma \ref{l1}, we can formulate conditions, when the contour $\mathbb{R}$ separates the <<upper>> series from the <<lower>> ones, as follows:
\begin{equation}
\label{cond}
\mathrm{Re}(\overline{\beta}_i)+n_{\beta_i}/2>0,\,\,\,\,\,\,i=2,3,\,\,\,\,\,\,
\mbox{and}\,\,\,\,\,\,
\mathrm{Re}(\overline{\beta}_j)+n_{\beta_j}/2>0,\,\,\,\,\,\,j=4,5,6.
\end{equation}

\noindent\textbf{Remark:} We will assume that conditions (\ref{cond}) are satisfied and the contour along the real axis separates the series of poles. All other cases we understand in the sense of analytic continuation. 

\begin{thm}
\label{th}
Taking into account the above assumptions and conditions
(\ref{cond}), integral (\ref{int1}) is equal to
\begin{equation*}
\pi^2D^{\alpha_{12345}-2}(z_2,z_1)(I_1+I_2+I_3),
\end{equation*}
where the explicit form for $I_i$, $i=1,2,3$, is given after relation (\ref{eq0}).
\end{thm}
\noindent\textbf{Proof:}
Using definition (\ref{function}), we represent the integrand as a product of Gamma functions. Next, we denote the product of factors containing the variable $s$ by $g(s)$, and the product with the variable $\overline{s}$ by $f(\overline{s})$.

Note that we will close the contour in the upper half-plane. Therefore, it is convenient to extract the singular parts of the functions and represent $f(\overline{s})$ in the form
\begin{equation}
f(\overline{s})=
\Gamma(\overline{s})f_1(\overline{s})=
\Gamma(\overline{\beta}_2+\overline{s})f_2(\overline{s})=
\Gamma(\overline{\beta}_3+\overline{s})f_3(\overline{s}).
\end{equation}

Then the integrand can be represented in three ways 
\begin{equation}
(-1)^n\frac{\Gamma(\overline{s})f_1(\overline{s})}{g(s)}=
(-1)^n\frac{\Gamma(\overline{\beta}_2+\overline{s})f_2(\overline{s})}{g(s)}=
(-1)^n\frac{\Gamma(\overline{\beta}_3+\overline{s})f_3(\overline{s})}{g(s)}.
\end{equation}

Using Lemma~\ref{l1}, we write down the series of poles with respect to variable $\nu$:

\begin{equation*}
\left\{i\left(\frac{
	\mid n_s\mid}{2}+k\right),\,k\in\mathbb{N}
\cup\{0\}\right\},
\end{equation*}
\begin{equation*}
\left\{i\left(\overline{\beta}_2+\frac{n_{\beta_2}+
	\mid n_s+n_{\beta_2}\mid}{2}+k\right),\,k\in\mathbb{N}
\cup\{0\}\right\},
\end{equation*}
\begin{equation*}
\left\{i\left(\overline{\beta}_3+\frac{n_{\beta_3}+
	\mid n_s+n_{\beta_3}\mid}{2}+k\right),\,k\in\mathbb{N}
\cup\{0\}\right\}.
\end{equation*}

Calculating the residues, we get three terms, each of which contains a double sum over the set  $\mathbb{Z}\times\mathbb{N}$. It turns out that we can transform the result, using the following statement 
\begin{lemma}
\label{l2}
Let $k\in\mathbb{N}$, $n,m\in\mathbb{Z}$, and $h_1,h_2\in C(\mathbb{Z},\mathbb{C})$ are functions that decrease sufficiently rapidly at infinity, then the following relation holds
\begin{multline*}
\sum\limits_{n\in\mathbb{Z}}\sum\limits_{k=0}^{\infty}(-1)^n
h_1\left(\frac{m+n+\mid n+m\mid}{2}+k\right)h_2\left(\frac{n-m-\mid n+m\mid}{2}-k\right)=\\
=
\left(\sum\limits_{k=0}^{\infty}(-1)^kh_1(k)\right)
\left(\sum\limits_{n=0}^{\infty}(-1)^{m+n}h_2(-m-n)\right).
\end{multline*}
\end{lemma}
\noindent\textbf{Proof:} Let us divide the summation over the variable $n\in\mathbb{Z}$ into two parts: $n+m\geqslant0$ and $n+m<0$.

We make the change of variables $n\to-m-k-n$, and then use the discrete Tonelli--Fubini theorem
\begin{equation*}
\sum\limits_{k=0}^{\infty}\sum\limits_{n=k}^{\infty}\to
\sum\limits_{n=0}^{\infty}\sum\limits_{k=0}^{n}.
\end{equation*}

In the second part of the sum we also make the change of variables $n\to-m-k+n$, then rename the variables $n\leftrightarrow k$, which gives the sum in the form
\begin{equation*}
\sum\limits_{n=0}^{\infty}\sum\limits_{k=n+1}^{\infty}.
\end{equation*}

Taking into account the fact that in both cases the summand function has the same form, the two expressions can be combined into one, which completes the proof of the Lemma. $\blacksquare$

Using Lemma \ref{l2}, we transform the result for each series of the poles:
\begin{equation*}
2\pi\sum\limits_{n_s\in\mathbb{Z}}\sum\limits_{k=0}^{\infty}
\frac{(-1)^{\left(\frac{n_s+\mid n_s\mid}{2}+k+n_s\right)}}{\left(\frac{n_s+\mid n_s\mid}{2}+k\right)!}
\frac{f_1\left(-\frac{n_s+\mid n_s\mid}{2}-k\right)}{g\left(\frac{n_s-\mid n_s\mid}{2}-k\right)}=2\pi
\left(\sum\limits_{k=0}^{\infty}\frac{f_1(-k)}{k!}\right)
\left(\sum\limits_{n=0}^{\infty}\frac{(-1)^{n}}{g(-n)}\right).
\end{equation*}

Similarly, we get the result for the two remaining series:
\begin{equation*}
2\pi
\left(\sum\limits_{k=0}^{\infty}\frac{f_2(-k-\overline{\beta}_2)}{k!}\right)
\left(\sum\limits_{n=0}^{\infty}\frac{(-1)^{n+n_{\beta_2}}}{g(-n-\beta_2)}\right),\,\,\,
2\pi
\left(\sum\limits_{k=0}^{\infty}\frac{f_3(-k-\overline{\beta}_3)}{k!}\right)
\left(\sum\limits_{n=0}^{\infty}\frac{(-1)^{n+n_{\beta_3}}}{g(-n-\beta_3)}\right).
\end{equation*}

The last sums can be transformed into hypergeometric functions, using the expansions 
\begin{multline*}
{}_3F_2(a_1,a_2,a_3;1-a_4,1-a_5;x)\prod\limits_{i=1}^{5}\Gamma(a_i)=
\\=
\sum\limits_{k=0}^{\infty}
\frac{x^k}{\Gamma(k+1)}
\Gamma(a_1+k)\Gamma(a_2+k)\Gamma(a_3+k)\Gamma(a_4-k)\Gamma(a_5-k),
\end{multline*}
\begin{multline*}
{}_3F_2(1-a_1,1-a_2,1-a_3;a_4,a_5;x)=
\\=\prod\limits_{i=1}^{5}\Gamma(a_i)
\sum\limits_{k=0}^{\infty}
\frac{(-x)^k}{\Gamma(k+1)
	\Gamma(a_1-k)\Gamma(a_2-k)\Gamma(a_3-k)\Gamma(a_4+k)\Gamma(a_5+k)}.
\end{multline*}

Then, taking into account the coefficient in front of the integral (\ref{int}) and the property  $D^{\alpha}(z_1,z_2)=(-1)^{n_{\alpha}}D^{\alpha}(z_2,z_1)$, we can write the final answer in the form of three terms:
\begin{equation}
\label{eq0}
\pi^2D^{\alpha_{12345}-2}(z_2,z_1)(I_1+I_2+I_3),
\end{equation}
where
\begin{align*}
&I_1=a(\alpha_1,\alpha_3,\alpha_4,2-\alpha_{14},
\alpha_{25},2-\alpha_{235})\\&
{}_3F_2(\alpha_5,1-\alpha_1,-1+\alpha_{235},
2-\alpha_{14},\alpha_{25};1)\,
{}_3F_2(\overline{\alpha_5},1-\overline{\alpha_1},
-1+\overline{\alpha_{235}},2-\overline{\alpha_{14}},
\overline{\alpha_{25}};1)\,,
\end{align*}
\begin{align*}
&I_2= a(\alpha_2,\alpha_4,\alpha_5,-1+\alpha_{125},
2-\alpha_{25},3-\alpha_{1245})
(-1)^{n_{\alpha_{23}}}\\&
{}_3F_2(\alpha_3,2-\alpha_{125},1-\alpha_2,2-\alpha_{25},
3-\alpha_{1245};1)\,
{}_3F_2(\overline{\alpha_3},
2-\overline{\alpha_{125}},1-\overline{\alpha_2},
2-\overline{\alpha_{25}},
3-\overline{\alpha_{1245}};1),
\end{align*}
\begin{align*}
&I_3=a(\alpha_3,\alpha_5,3-\alpha_{12345},2-\alpha_{145},\alpha_{14},
-1+\alpha_{1245})(-1)^{n_{\alpha_{15}}}\\&
{}_3F_2(\alpha_4,-2+\alpha_{12345},
-1+\alpha_{145},\alpha_{14},
-1+\alpha_{1245};1)
{}_3F_2(\overline{\alpha_4},
-2+\overline{\alpha_{12345}},
-1+\overline{\alpha_{145}},
\overline{\alpha_{14}},-1+\overline{\alpha_{1245}};1)\,,
\end{align*}
which completes the proof of Theorem \ref{th}. $\blacksquare$\\

\section{Multidimensional case}

In this case family of functions~(\ref{family}) has the form
\begin{equation}\label{dfamily}
D^{\alpha}(z_1,z_2) = \dfrac{(z_1-z_2)^{\mu_1\ldots\mu_n}}{(z_1-z_2)^{2\left(d/4 + n/2 + i\nu\right)}},
\end{equation}
where $\nu \in \mathbb{R} \ \,,\ n=0,1,2,\ldots$, and $z_1\,,z_2$ are vectors in $d$-dimensional space ($z_1\,,z_2 \in \mathbb{R}^d$).
The symbol $(z_1-z_2)^{\mu_1\ldots\mu_n}$ means the traceless part of the tensor
$(z_1-z_2)^{\mu_1}\cdots(z_1-z_2)^{\mu_n}$, for example
\begin{align*}
(z_1-z_2)^{\mu_1\mu_2} = (z_1-z_2)^{\mu_1}(z_1-z_2)^{\mu_2} -
\dfrac{1}{d}\delta^{\mu_1\mu_2}(z_1-z_2)^2.
\end{align*}

In the general case we have expressions (\ref{P}) and (\ref{P1}) as explicit formulae.
Thus, $\alpha$ has discrete parameters and one real continuous parameter $\nu$, as we had in the two-dimensional case. Discrete parameters include the number $n$,
which specifies the rank of the tensor $(z_1-z_2)^{\mu_1\ldots\mu_n}$, and the number that specifies independent components of the symmetric traceless tensor of rank $n$. We have only two independent components in the two dimensional case, so only $n$ remains.
We integrate over region $\mathcal{V}$, which coincides with  $\mathbb{R}^d$, and the integration measure has the standard form $\mathcal{D}z=d^dz$. We will omit notion of the integration region $\mathbb{R}^d$ in most situations.

The chain rule, shown in Fig. \ref{chain}, gives the answer for the integral
\begin{equation}\label{dchain1}
\int d^{d} x \dfrac{x^{\mu_1\ldots\mu_n}}{x^{2\alpha}}\dfrac{(y-x)^{\mu_{n+1}\ldots\mu_{n+m}}}{\left(y - x\right)^{2\beta}},
\end{equation}
where $n,m\in\mathbb{N}$. 
But we do not need the general formula to analyze the considered diagram, therefore, we are going to use only some special cases.

In particular, if we put $m = 0$ in formula~(\ref{dchain1}), we can obtain the following chain relation~\cite{AN,KT}
\begin{equation}\label{dchain}
\int d^{d} x \dfrac{x^{\mu_1\ldots\mu_n}}{x^{2\alpha}}\dfrac{1}{\left(y - x\right)^{2\beta}} = \pi^{d/2}\dfrac{a_n\left(\alpha\right)a_0\left(\beta\right)}{a_n{\left(\alpha + \beta - d/2\right)}}\dfrac{y^{\mu_1\ldots\mu_n}}{y^{2\left(\alpha + \beta - d/2\right)}},
\end{equation}
where
\begin{equation*}
a_n\left(\alpha\right) = \dfrac{\Gamma\left(d/2 - \alpha + n\right)}{\Gamma\left(\alpha\right)}\,.
\end{equation*}

We also need one more relation. Setting $m = n$ in (\ref{dchain1}), and contracting $\mu_i$ with $\mu_{n+i}$ for $i\in\{1,\ldots,n\}$, we get an integral, that has the following form after calculation
\begin{align}\label{tchain}
\int d^{d}x\dfrac{x^{\mu_1\ldots\mu_n}}{x^{2\alpha}}\dfrac{\left(y - x\right)^{\mu_1\ldots\mu_n}}{\left(y - x\right)^{2\beta}} = (-1)^n\dfrac{\pi^{d/2}a_n\left(\alpha\right)a_n\left(\beta\right)}{a_0\left(\alpha + \beta - d/2 - n\right)}\dfrac{\Gamma\left(n + d - 2\right)\Gamma\left(d/2 - 1\right)}{2^n\Gamma\left(d - 2\right)\Gamma\left(n + d/2 - 1\right)}\dfrac{1}{y^{2\left(\alpha + \beta - d/2 - n\right)}},
\end{align}
where summation over the repeated indices is implied. Let us note that the l.h.s of  (\ref{tchain}) contains integrals of the form (\ref{dchain1}) by construction, so the coefficient in the r.h.s is given by the sum of the same $B$-coefficients shown in Fig. \ref{chain}. Thus, we do not need distinct $B$-coefficients, but only their sum, which can be calculated in a closed form.  

Now let us proceed to the derivation of the Mellin-Barnes representation, shown in Fig. \ref{mb}.
Functions of the form~(\ref{dfamily}), which are the eigenfunctions of the commuting operators $Q\left(u\right)$ form a complete orthogonal system (see Appendix).
The orthogonality and completeness relations have the following explicit forms
\begin{equation}
\int d^d x \dfrac{x^{\mu_1\ldots\mu_n}}{x^{2\left(d/4 + n/2 + i\nu\right)}}\dfrac{x^{\nu_1\ldots\nu_m}}{x^{2\left(d/4 + m/2 - i\lambda\right)}} = c_n\delta_{nm}\delta\left(\nu - \lambda\right)P^{\mu_1\ldots\mu_n}_{\nu_1\ldots\nu_n},
\end{equation}
\begin{equation}\label{compl}
\sum_{n \ge 0}\dfrac{1}{c_n}\int_\mathbb{R}
d\nu \dfrac{x^{\mu_1\ldots\mu_n}}{x^{2\left(d/4 + n/2 + i\nu\right)}}\dfrac{y^{\mu_1\ldots\mu_n}}{y^{2\left(d/4 + n/2 - i\nu\right)}} = \delta^{\left(d\right)}\left(x - y\right),
\end{equation}
where the constant $c_n$ is given by the expression
\begin{equation}
\label{coef}
c_n = \dfrac{\pi^{d/2 + 1}n!}{2^{n - 1}\Gamma\left(d/2 + n\right)},
\end{equation}
and summation over the repeated indices in~(\ref{compl}) is implied.

Operator $P^{\mu_1\ldots\mu_n}_{\nu_1\ldots\nu_n}$ is a projector onto the traceless symmetric tensors
\begin{equation}\label{P}
x^{\mu_1\ldots\mu_n} = P^{\mu_1\ldots\mu_n}_{\nu_1\ldots\nu_n}x^{\nu_1}\ldots x^{\nu_n}
\end{equation}
and given by the formula~\cite{K,KT}
\begin{equation}\label{P1}
P^{\mu_1\ldots\mu_n}_{\nu_1\ldots\nu_n} = \sum_{p \ge 0}\dfrac{n!\left(-1\right)^{p}\Gamma\left(n - p + d/2 - 1\right)}{2^{2p}p!\left(n - 2p\right)!\Gamma\left(n + d/2 - 1\right)}\hat{S}\left(\delta^{\mu_1\mu_2}\delta_{\nu_1\nu_2}\ldots\delta^{\mu_{2p -1}\mu_{2p}}\delta_{\nu_{2p-1}\nu_{2p}}\delta^{\mu_{2p + 1}}_{\nu_{2p + 1}\ldots}\delta^{\mu_{n}}_{\nu_{n}}\right),
\end{equation}
where $\hat{S}$ denotes the symmetrization over all indices.
Using the completeness relation, we can get the expression for the scalar propagator in the Mellin-Barnes representation. The simplest way is to expand the $\delta$-function in
\begin{equation*}
\dfrac{1}{\left(x - y\right)^{2\alpha}} = \int d^{d}z \dfrac{1}{\left(x - z\right)^{2\alpha}}\,\delta^{\left(d\right)}\left(z - y\right),
\end{equation*}
using~\eqref{compl}, which gives
\begin{align*}
\dfrac{1}{\left(x - y\right)^{2\alpha}} = \sum_{n\geq 0}\int_{\mathbb{R}} d\nu \dfrac{1}{c_n}\dfrac{y^{\mu_1\ldots\mu_n}}{y^{2\left(d/4 + n/2 - i\nu\right)}}\int d^d z \dfrac{1}{\left(z - y\right)^{2\alpha}}\dfrac{z^{\mu_1\ldots\mu_n}}{z^{2\left(d/4 + n/2 + i\nu\right)}}.
\end{align*}

To calculate the integral with respect to $z$ we use chain rule~(\ref{dchain}), thus obtaining a representation for the propagator
\begin{align*}
\dfrac{1}{\left(x - y\right)^{2\alpha}} = \sum_{n\geq 0}\int_{\mathbb{R}} d\nu \dfrac{\pi^{d/2}}{c_n}\dfrac{a_n\left(d/4 + n/2 + i\nu\right)a_0\left(\alpha\right)}{a_n\left(\alpha - d/4 + n/2 + i\nu\right)}\dfrac{x^{\mu_1\ldots\mu_n}}{x^{2\left(\alpha - d/4 + n/2 + i\nu\right)}}\dfrac{y^{\mu_1\ldots\mu_n}}{y^{2\left(d/4 + n/2 - i\nu\right)}}.
\end{align*}

The last expression is invariant under translation by an arbitrary vector, so that the general formula for the Mellin-Barnes transform has the following form  
\begin{align}\label{MBD}
\dfrac{1}{\left(x - y\right)^{2\alpha}} = \sum_{n\geq 0}\int_{\mathbb{R}} d\nu \dfrac{\pi^{d/2}}{c_n}\dfrac{a_n\left(d/4 + n/2 + i\nu\right)a_0\left(\alpha\right)}{a_n\left(\alpha - d/4 + n/2 + i\nu\right)}\dfrac{\left(x - z\right)^{\mu_1\ldots\mu_n}}{\left(x - z\right)^{2\left(\alpha - d/4 + n/2 + i\nu\right)}}\dfrac{\left(y - z\right)^{\mu_1\ldots\mu_n}}{\left(y - z\right)^{2\left(d/4 + n/2 - i\nu\right)}}.
\end{align}

To avoid confusion, let us note that summation with respect to the Mellin--Barnes measure is done over $n$ as well as the tensor indices $\mu_i$, $i\in\{1,\ldots,n\}$.

Now let us compute the diagram shown in Fig. \ref{dia}. We do not have the formula for arbitrary parameters in (\ref{dchain1}) and we can not compute the diagram in the most general form. For this reason, we consider functions (\ref{dfamily}) only in scalar form, that is, when $n=0$. It turns out that in this case, precisely due to relation (\ref{tchain}), it becomes possible to calculate the integral (\ref{int}).

The two-loop integral
\begin{equation}
I\left(\alpha_1,\alpha_2,\alpha_3,\alpha_4,\alpha_5\right) = \int \dfrac{d^d x d^d y}{x^{2\alpha_1}y^{2\alpha_2}\left(z - x\right)^{2\alpha_4}\left(z - y\right)^{2\alpha_3}\left(x - y\right)^{2\alpha_5}}
\end{equation}
is computed, according to the general scheme, outlined in the second section.
At the first step we use formula ~(\ref{MBD}) for the propagator between vertices $x$ and $z$, shifting both vertices by the vector $y$.
The integral takes the form
\begin{align}
\label{dia1}
I\left(\alpha_1,\alpha_2,\alpha_3,\alpha_4,\alpha_5\right) =
&\sum_{n\geq 0}\int_{\mathbb{R}} d\nu \dfrac{\pi^{d/2}}{c_n}\dfrac{a_n\left(d/4 + n/2 + i\nu\right)a_0\left(\alpha_4\right)}{a_n\left(\alpha_4 + n/2 - d/4 + i\nu\right)} \\\nonumber &\int d^d x d^d y \dfrac{\left(x - y\right)^{\mu_1\ldots\mu_n}\left(z - y\right)^{\mu_1\ldots\mu_n}}{x^{2\alpha_1}y^{2\alpha_2}\left(x - y\right)^{2\left(\alpha_{45} - d/4 + n/2  + i\nu\right)}\left(z - y\right)^{2\left(\alpha_3 + d/4 + n/2 - i\nu\right)}},
\end{align}
so we can compute the remaining integrals over $x$ and $y$ by using the chain rule.

Indeed, let us use relation (\ref{dchain}) and calculate integral over $x$. Then, the second line of (\ref{dia1}) can be rewritten as
\begin{equation*}
\left(-1\right)^n \pi^{d/2}\dfrac{a_n\left(\alpha_{45} - d/4 + n/2  + i\nu\right)a_0\left(\alpha_1\right)}{a_n\left(\alpha_{145} - 3d/4 + n/2 + i\nu\right)}
\int d^dy \dfrac{y^{\mu_1\ldots\mu_n}}{y^{\alpha_{1245} - 3d/4 + n/2 + i\nu}}\dfrac{\left(z - y\right)^{\mu_1\ldots\mu_n}}{\left(z - y\right)^{2\left(\alpha_3 + d/4 + n/2 - i\nu\right)}}.
\end{equation*}

Next, according to the scheme from Section 2, we need to apply the chain rule to every summand in the sum over tensor indices. However, due to the absence of the needed chain relation, we add summation over tensor indices to the integration over the variable $y$, and then use relation~(\ref{tchain}). Then we get equality
\begin{align*}
\int d^dy \dfrac{y^{\mu_1\ldots\mu_n}}{y^{\alpha_{1245} - 3d/4 + n/2  + i\nu}}\dfrac{\left(z - y\right)^{\mu_1\ldots\mu_n}}{\left(z - y\right)^{2\left(\alpha_3 + d/4 + n/2 - i\nu\right)}} = (-1)^n\pi^{d/2}\dfrac{\Gamma\left(n + d - 2\right)\Gamma\left(d/2 - 1\right)}{2^n\Gamma\left(d - 2\right)\Gamma\left(n + d/2 - 1\right)}\\ \dfrac{a_n\left(\alpha_{1245} - 3d/4 + n/2 + i\nu\right)a_n\left(\alpha_3 + d/4 + n/2 - i\nu\right)}{a_0\left(\alpha_{12345} - d\right)}\dfrac{1}{z^{2\left(\alpha_{12345} - d\right)}}.
\end{align*}

Thus, the result for the diagram has the following general form
\begin{equation*}
I\left(\alpha_1,\alpha_2,\alpha_3,\alpha_4,\alpha_5\right) = \dfrac{C}{z^{2\left(\alpha_{12345} - d\right)}},
\end{equation*}
where
\begin{align}
C = &\dfrac{\pi^{d - 1}}{2}\dfrac{\Gamma\left(d/2 - 1\right)}{\Gamma\left(d - 2\right)}\dfrac{a_0\left(\alpha_1\right)a_0\left(\alpha_4\right)}
{a_0\left(\alpha_{12345} - d\right)}\sum_n\int_{\mathbb{R}} d\nu \,M_n\,
\dfrac{a_n\left(d/4 + n/2 + i\nu\right)}{a_n\left(\alpha_4 - d/4  + n/2+ i\nu\right)}\\
\nonumber
&\dfrac{a_n\left(\alpha_{45} - d/4 + n/2 + i\nu\right)}{a_n\left(\alpha_{145} - 3d/4 + n/2  + i\nu\right)}a_n\left(\alpha_{1245}- 3d/4 + n/2  + i\nu\right)
a_n\left(\alpha_3+ d/4 + n/2 - i\nu\right),
\end{align}
where we use the convenient notation
\begin{equation}
M_n = \dfrac{1}{n!}\left(n+d/2 -1\right)\Gamma\left(n + d - 2\right).
\end{equation}

We calculate the integral over $\nu$ by residues by analogy with the two-dimensional case. In order to understand the structure of the poles, we rewrite the integrand in terms of Gamma functions
\begin{align*}
\dfrac{\Gamma\left(d/4 + n/2 - i\nu\right)}{\Gamma\left(d/4 + n/2 + i\nu\right)}&\dfrac{\Gamma\left(\alpha_4 - d/4 + n/2 + i\nu\right)}{\Gamma\left(3d/4 - \alpha_4 + n/2 - i\nu\right)}\dfrac{\Gamma\left(3d/4 - \alpha_{45} + n/2  - i\nu\right)}{\Gamma\left(\alpha_{45} - d/4 + n/2 + i\nu\right)}\dfrac{\Gamma\left(\alpha_{145}- 3d/4 + n/2 + i\nu\right)}{\Gamma\left(5d/4- \alpha_{145} + n/2  - i\nu\right)}\\
&\dfrac{\Gamma\left(5d/4 - \alpha_{1245} + n/2 - i\nu\right)}{\Gamma\left(\alpha_{1245} - 3d/4 + n/2 + i\nu\right)}\dfrac{\Gamma\left(d/4 - \alpha_3 + n/2 + i\nu\right)}{\Gamma\left(\alpha_3 + d/4 + n/2 - i\nu\right)}.
\end{align*}

This expression has six infinite series of poles, three of them are located in the upper half plane and three in the lower one. As in the two-dimensional case, we will close the contour in the upper half-plane, so we proceed to enumerating the corresponding poles. The gamma function has simple poles at the non-positive integers.
\begin{equation*}
\Gamma\left(x\right) = \dfrac{\left(-1\right)^{n}}{n!}\dfrac{1}{x+n} + \ldots,
\end{equation*}
so that the poles in the upper half-plane appear in three cases 
\begin{equation*}
\begin{cases}
\nu = i\left(k_1 + \alpha_4 - d/4 + n/2\right);\ \\
\nu = i\left(k_2 + \alpha_1 + \alpha_4 + \alpha_5 - 3d/4 + n/2\right);\ \\
\nu = i\left(k_3 + d/4 - \alpha_3 + n/2\right).
\end{cases}
\end{equation*}

Summing the residues, we get the final answer
\begin{equation}
C = \pi^d \dfrac{\Gamma\left(d/2 - 1\right)}{\Gamma\left(d - 2\right)}\dfrac{a_0\left(\alpha_1\right)a_0\left(\alpha_4\right)}
{a_0\left(\alpha_{12345} - d\right)}\,\left(I_1 + I_2 + I_3\right),
\end{equation}
where
\begin{align*}
I_1 = \sum_{n = 0}^{\infty}M_n\sum_{k = 0}^{\infty}\dfrac{\left(-1\right)^k}{k!}\dfrac{\Gamma\left(\alpha_4 + n + k\right)}{\Gamma\left(d/2 - \alpha_4 - k\right)}\dfrac{\Gamma\left(d/2- \alpha_5 + n + k\right)}{\Gamma\left(\alpha_5 - k\right)}\dfrac{\Gamma\left(\alpha_{15} - d/2 - k\right)}{\Gamma\left(d - \alpha_{15} + n + k\right)}\\ \nonumber
\dfrac{\Gamma\left(d - \alpha_{125} + n  + k\right)}{\Gamma\left(\alpha_{125} - d/2 - k\right)}\dfrac{\Gamma\left(d/2 - \alpha_{34} - k\right)}{\Gamma\left( \alpha_{34} + n + k\right)}\dfrac{1}{\Gamma\left(d/2 + n + k\right)},
\end{align*}
\begin{align*}
I_2 = \sum_{n = 0}^{\infty}M_n\sum_{k = 0}^{\infty}\dfrac{\left(-1\right)^k}{k!}\dfrac{\Gamma\left(\alpha_{145} - d/2 + n 
 + k\right)}{\Gamma\left(d - \alpha_{145} - k\right)}\dfrac{\Gamma\left(d/2 - \alpha_{15} - k\right)}
{\Gamma\left(\alpha_{15} + n + k\right)}
\dfrac{\Gamma\left(\alpha_1 + n + k\right)}
{\Gamma\left(d/2 - \alpha_1 - k\right)}\\ \nonumber
\dfrac{\Gamma\left(d/2 - \alpha_2 + n + k\right)}{\Gamma\left(\alpha_2 - k\right)}\dfrac{\Gamma\left(d - \alpha_{1345} - k\right)}{
\Gamma\left(\alpha_{1345} - d/2 + n + k\right)}\dfrac{1}{\Gamma\left(d/2 + n + k\right)},
\end{align*}
\begin{align*}
I_3 = \sum_{n = 0}^{\infty}M_n\sum_{k = 0}^{\infty}\dfrac{\left(-1\right)^k}{k!}\dfrac{\Gamma\left(d/2 - \alpha_3 + n + k\right)}{\Gamma\left(\alpha_3 - k\right)}\dfrac{\Gamma\left(\alpha_{34} - d/2 - k\right)}{\Gamma\left(d - \alpha_{34} + n + k\right)}
\dfrac{\Gamma\left(d - \alpha_{345} + n + k\right)}
{\Gamma\left(\alpha_{345} - d/2 - k\right)}\\ \nonumber
\dfrac{\Gamma\left(\alpha_{1345} - d - k\right)}{\Gamma\left(3d/2 - \alpha_{1345} + n + k\right)}\dfrac{\Gamma\left(3d/2 - \alpha_{12345} + n + k\right)}{\Gamma\left(\alpha_{12345}- d - k\right)}\dfrac{1}{\Gamma\left(d/2 + n + k\right)}.
\end{align*}

\section{Conclusion}

We have covered in detail the application of the simplest integral transform to the separated variables representation to compute the two-loop master-diagram. Apparently, the proposed approach is equivalent to the Gegenbauer polynomial technique~\cite{ChKT,K,KT,Sch}, so our work is rather methodological.

The next step is to compare the obtained formulae with the numerous already available results~\cite{AN,Gr,BW,KT}. In particular, an important task is to calculate a special master diagram, arising in the $1/N$ expansion~\cite{AN,VPH,BK}. 

The proposed method can be applied to the calculation of ladder diagrams~\cite{UD,B,Isa}. In the multidimens\-ional case all calculations repeat step by step those in the two-dimensional~\cite{DKO} and four-dimensional~\cite{KO} cases.
The corresponding formulae are given in the Appendix.

Let us note interesting connections with other works. The two-dimensional transformation to separated variables turned out to be a convenient tool in the study of Racah operators in the representation theory of $SL(2, \mathbb{C})$, to which the works of R.S. Ismagilov~\cite{I1,I2} and~\cite{DS} are devoted. The works of Yu.A. Neretin~\cite{N1,N2} explore the Barnes--Ismagilov integrals and their generalizations, and the work of Yu.A. Neretin and V.F. Molchanov~\cite{MN} essentially investigated in detail the second transformation from the hierarchy of transformations to separated variables. 

The work~\cite{DMV} is devoted to the application of the transform to separated variables to derive a two-dimensional version of the Gustafson integrals. An interesting open question is the study of the connection between the method of separation of variables and the Dotsenko--Fateev and Fateev--Litvinov integral identities in conformal field theory\cite{DF,FL}.

\paragraph{Acknowledgments:}

This work was supported by the Russian Science Foundation (project 19-11-00131). Also A.V. Ivanov is a winner of the Young Russian Mathematician award and would like
to thank its sponsors and jury.

\section{Appendix}
%

\subsection{Projector}

Let us introduce a projector onto the space of symmetric and traceless tensors
\begin{equation*}
\psi^{\prime}_{\mu_1\ldots\mu_n} = P^{\mu_1\ldots\mu_n}_{\nu_1\ldots\nu_n}\,
\psi_{\nu_1\ldots\nu_n}.
\end{equation*}

The initial tensor $\psi_{\nu_1\ldots\nu_n}$ is symmetric and its image $\psi^{\prime}_{\mu_1\ldots\mu_n}$ is traceless. Considering a polynomial of the form
\begin{equation*}
    P\left(u, v\right) = \dfrac{1}{n!} P^{\mu_1\ldots\mu_n}_{\nu_1\ldots\nu_n}u^{\mu_1}\cdots u^{\mu_n}v^{\nu_1}\cdots v^{\nu_n}
\end{equation*}
and rewriting previous equality by using the generating function
\begin{equation*}
\psi^{\prime}\left(u\right) = \psi^{\prime}_{\mu_1\ldots\mu_n}u^{\mu_1}\ldots u^{\mu_n},\,\,\,\,  \psi\left(v\right) = \psi_{\mu_1\ldots\mu_n}v^{\mu_1}\ldots v^{\mu_n},
\end{equation*}
we get
\begin{equation*}
\psi^{\prime}\left(u\right) = P\left(u, \partial_{v}\right)\psi\left(v\right).
\end{equation*}

The tracelessness of
$\psi^{\prime}_{\mu_1\ldots\mu_n}$ is equivalent to $\Delta_u P(u,v) = 0$. Substituting the general decomposition
\begin{equation*}
P\left(u,v\right) = \left(u v\right)^n + a_1\left(u v\right)^{n - 2}u^2v^2 + a_2\left(u v\right)^{n - 4}u^4v^4 + \ldots
\end{equation*}
in the equation $\Delta_u P(u,v) = 0$, we obtain the recurrence relation
\begin{equation}
    a_p = -a_{p - 1}\dfrac{\left(n - 2p + 2\right)\left(n - 2p + 1\right)}{2p\left(d + 2n - 2 - 2p\right)}\,.
\end{equation}
A solution which satisfies the initial condition $a_0=1$ has the form
\begin{equation}
    a_p = \dfrac{n!\left(-1\right)^{p}\Gamma\left(n - p + d/2 - 1\right)}{2^{2p}p!\left(n - 2p\right)!\Gamma\left(n + d/2 - 1\right)}\,.
\end{equation}
Thus, we have formula (\ref{P1}),
where $\hat{S}$ denotes the symmetrization over all indices
\begin{align}
\hat{S}\,\psi_{\mu_1\cdots\mu_n} =
\frac{1}{n!} \sum_{p\in S_n} \psi_{\mu_{p(1)}\cdots\mu_{p(n)}}.
\end{align}

\subsection{Orthogonality}

The orthogonality relation has the following form
\begin{equation}
\int d^d x \dfrac{x^{\mu_1\ldots\mu_n}}{x^{2\left(d/4 + n/2 + i\nu\right)}}\dfrac{x^{\nu_1\ldots\nu_m}}{x^{2\left(d/4 + m/2 - i\lambda\right)}} = c_n\delta_{nm}\delta\left(\nu - \lambda\right)P^{\mu_1\ldots\mu_n}_{\nu_1\ldots\nu_n}.
\end{equation}

We obtain $\delta\left(\nu - \lambda\right)$ by integrating over the radial variable, therefore it is convenient to separate this part by going to the spherical coordinates 
\begin{equation*}
\int d^d x \dfrac{x^{\mu_1\ldots\mu_n}\,x^{\nu_1\ldots\nu_m}}
{x^{2\left(d/2 + (n+m)/2 + i(\nu-\lambda)\right)}} =
\pi\delta(\lambda-\nu)\,\int_{\mathbb{S}^{d-1}} d \Omega\, \hat{x}^{\mu_1\ldots\mu_n}\,\hat{x}^{\nu_1\ldots\nu_m},
\end{equation*}
where $d\Omega$ is the standard measure on the sphere,
$\hat{x}^{\mu}$ is a vector of unit length, and the integral over the radius is done by substituting $r = e^p$
\begin{equation*}
\int^{\infty}_{0} \dfrac{dr}{r} r^{2i(\lambda-\nu)} = \pi\delta(\lambda-\nu).
\end{equation*}

Thus, we need to prove the identity
\begin{align*}
\int_{\mathbb{S}^{d-1}} d \Omega\,
\hat{x}^{\mu_1\ldots\mu_n}\,
\hat{x}^{\nu_1\ldots\nu_m} = \dfrac{c_n}{\pi}\,\delta_{nm}\,P^{\mu_1\ldots\mu_n}_{\nu_1\ldots\nu_n}.
\end{align*}

Consider an equivalent relation by contracting both sides with complex vectors $u$ and $v$, satisfying the condition $u^2=v^2=0$,
\begin{align*}
\int_{\mathbb{S}^{d-1}} d \Omega
\left(\hat{x}u\right)^{n}\,\left(\hat{x}v\right)^{m} = \dfrac{c_n}{\pi}\,\delta_{nm}\,\left(uv\right)^{n},
\end{align*}
and prove this relation by using the standard formula for the Gaussian integral
\begin{align*}
\int d^dx\,e^{-x^2+xa} = \pi^{d/2} e^{\frac{a^2}{4}}.
\end{align*}

Substituting $a = t u+s v$ and differentiating the 
required number of times by $t$ and $s$, we get
\begin{align*}
\int d^dx e^{-x^2}\,(xu)^n\,(xv)^m =
\pi^{d/2}n!2^{-n}\,\delta_{nm}(uv)^n.
\end{align*}

The integral above differs from the one we need by a constant factor which can be restored by going to the spherical coordinates

\begin{align*}
\int d^dx\,e^{-x^2}\,(xu)^n\,(xv)^m =
\frac{1}{2}\Gamma\left(\frac{d+n+m}{2}\right)\int_{\mathbb{S}^{d-1}} d \Omega\, (\hat{x}u)^n\,(\hat{x}v)^m,
\end{align*}
so that the constant $c_n$ has the form (\ref{coef}).

\subsection{Completeness}

The completeness relation has the form ($x, y \in \mathbb{R}^d$)
\begin{equation}\label{compl1}
    \sum_{n \ge 0}\int_{\mathbb{R}}d\nu\dfrac{1}{c_n}\dfrac{x^{\mu_1\ldots\mu_n}}{x^{2\left(d/4 + n/2 - i\nu\right)}}\dfrac{y^{\mu_1\ldots\mu_n}}{y^{2\left(d/4 + n/2 + i\nu\right)}} = \delta^{\left(d\right)}\left(x - y\right).
\end{equation}

We can expand the delta function in the r.h.s. as a product of the radial and angular parts
\begin{equation}
    \delta^{d}\left(x - y\right) = \dfrac{1}{x^{2\left(d/2 - 1/2\right)}}\delta\left(|x| - |y|\right)\dfrac{1}{J}\delta\left(\hat{x} - \hat{y}\right),
\end{equation}
where $|x|$ is the length of the vector $x$, $\hat{x} = x/|x|~$ is a vector of unit length, and $J$ is the Jacobian, corresponding to change of variables to $d$-dimensional spherical coordinates.

The l.h.s of~(\ref{compl1}) has a similar structure. 
The integral over $\nu$ depends only on the length of the vectors, and the sum over $n$ is related to the angular part. 
We can integrate over $\nu$ by substituting $|x| = e^u$ \,, $|y| = e^{v}$
\begin{equation}
    \int_{\mathbb{R}} d\nu \dfrac{1}{|x|^{-2i\nu}}\dfrac{1}{|y|^{2i\nu}} = \pi\delta\left(u - v\right).
\end{equation}
Therefore we need to prove the relation for the angular part only
\begin{equation}\label{ancompl}
    \sum_{n\ge 0}\dfrac{\pi}{c_n}\,\hat{x}^{\mu_1\ldots\mu_n}\hat{y}^{\mu_1\ldots\mu_n} = \dfrac{1}{J}\delta\left(\hat{x} - \hat{y}\right),
\end{equation}
which we can rewrite as the completeness relation for Gegenbauer polynomials
\begin{equation}\label{Gegen}
    \sum_{n\geq 0} \left(2n + d - 2\right) C_n^{\left(d/2 - 1\right)}\left(\hat{x}\cdot\hat{y}\right) =
    \left(d - 2\right)S_{d - 1}\,
    \dfrac{1}{J}\delta\left(\hat{x} - \hat{y}\right),
\end{equation}
where $S_{d-1} = 2\pi^{d/2}\Gamma^{-1}(d/2)$ is the surface of the unit sphere in $d$-dimensional space by using the relation between Gegenbauer polynomials and traceless tensors~\cite{ChKT}
\begin{equation}
	\label{geg}
    \hat{x}^{\mu_1\ldots\mu_n}\hat{y}^{\mu_1\ldots\mu_n} = \dfrac{n!\Gamma\left(d/2 - 1\right)}{2^n\Gamma\left(n + d/2 - 1\right)}C_{n}^{\left(d/2 - 1\right)}\left(\hat{x}\cdot\hat{y}\right),
\end{equation}

The function
\begin{equation*}
    G\left(x , y\right) = \dfrac{1}{\left(x- y\right)^{2\left(d/2 - 1\right)}}
\end{equation*}
is the fundamental solution of the $d-$ dimensional Laplace equation
\begin{equation}\label{laplace}
    -\operatorname{\Delta}_x G\left(x , y\right) = \left(d - 2\right)S_{d-1}\delta^{\left(d\right)}\left(x-y\right),
\end{equation}
and can be considered as a generating function for Gegenbauer polynomials~\cite{K}
(see multipole expansion)
\begin{equation}\label{mult1}
     G\left(x,y\right) = \begin{cases}
\sum_{n\ge 0}\dfrac{|x|^{n}}{|y|^{n + d - 2}}\,C^{\left(d/2 - 1\right)}_n\left(\hat{x}\cdot\hat{y}\right),\ |x|< |y|;\\
\sum_{n\ge 0}\dfrac{|y|^{n}}{|x|^{n + d - 2}}\,C^{\left(d/2 - 1\right)}_n\left(\hat{x}\cdot\hat{y}\right),\ |y|< |x|.
\end{cases}
\end{equation}

Integrating (\ref{laplace}) over $r =|x|$ in a neighborhood of the point $|y|$, we get
\begin{equation*}
    -r^{d - 1}\dfrac{\partial}{\partial r}G\left(x , y\right)\bigg|_{|y| - 0}^{|y| + 0} = \left(d - 2\right)\dfrac{S_{d-1}}{J}\delta\left(\hat{x} - \hat{y}\right).
\end{equation*}

If we substitute decomposition~(\ref{mult1}) into the l.h.s, we immediately obtain identity~(\ref{Gegen}).

\subsection{Operator $Q(u)$ and ladder diagrams}

Consider the integral operator $Q\left(u\right)$
\begin{equation*}
    \left[Q\left(u\right)\Phi\right]\left(x\right) = 
    \int d^d y\dfrac{1}{\left(x - y\right)^{2\left(d/2 - u\right)}y^{2u}}\Phi\left(y\right).
\end{equation*}

The operators $Q\left(u\right)$ form a commutative family. Commutativity 
$Q\left(u\right)Q\left(v\right) = Q\left(v\right)Q\left(u\right)$
can be proved by finding the kernel of the operator
$Q\left(u\right)Q\left(v\right)$, using the star-triangle relation~\cite{AN,Isa}, and checking that it is invariant under permutation $u\rightleftarrows v$.

We can check that the functions
\begin{equation*}
\Psi^{\mu_1\ldots\mu_n}\left(x\right) = \dfrac{x^{\mu_1\ldots\mu_n}}
{x^{2\left(d/4 + n/2  + i\nu\right)}}
\end{equation*}
are eigenfunctions of the operator $Q(u)$ explicitly by using chain rule~(\ref{dchain})
\begin{align*}
\left[Q\left(u\right)\Psi^{\mu_1\ldots\mu_n}\right]\left(x\right) =
\tau\left(u, \nu, n\right)\,
\Psi^{\mu_1\ldots\mu_n}(x),\\
\tau\left(u, \nu, n\right) = \pi^{d/2}\dfrac{a_n\left(d/4 + n/2 + u + i\nu\right)a_0\left(d/2 - u\right)}{a_n\left(d/4 + n/2 + i\nu\right)}.
\end{align*}

The system of functions $\Psi^{\mu_1\ldots\mu_n}\left(x\right)$ is orthogonal and complete, so that we obtain the spectral decomposition of for the kernel of operator $Q\left(u\right)$ in the form
\begin{align*}
Q_u\left(x, y\right) = 
    \dfrac{1}{\left(x - y\right)^{2\left(d/2 - u\right)}y^{2u}} = 
    \sum_{n \ge 0}\dfrac{1}{c_n}\int_{\mathbb{R}}d\nu~\tau\left(u, \nu, n\right)\dfrac{x^{\mu_1\ldots\mu_n}}{x^{2\left(d/4 + n/2  + i\nu\right)}}\dfrac{y^{\mu_1\ldots\mu_n}}{y^{2\left(d/4 + n/2 - i\nu\right)}}.
\end{align*}

Since the operators $Q(u)$ form a commutative family and have a common set of eigenfunctions, it follows that we can extend the previous formula to the product $Q(u_1)\cdots Q(u_{L})$ 
\begin{equation}\label{L}
Q_{u_1\ldots u_{L}}\left(x, y\right) = 
\sum_{n\ge 0}\dfrac{1}{c_n}\int_\mathbb{R}
d\nu\dfrac{x^{\mu_1\ldots\mu_n}}{x^{2\left(d/4 + n/2 + i\nu\right)}}\dfrac{y^{\mu_1\ldots\mu_n}}{y^{2\left(d/4 + n/2 - i\nu\right)}}\prod\limits_{k = 1}^{L}\tau\left(u_k,\nu,n\right).
\end{equation}

Now if we recall that the kernel of the product $Q(u_1)\cdots Q(u_{L})$ is given by the convolution of the kernels of $Q(u)$, we get
\begin{equation*}
    Q_{u_1\ldots u_{L}}\left(x, y\right) = 
    \prod\limits_{k = 1}^{L-1}\int d^d x_k \dfrac{1}{\left(x_{k - 1} - x_k\right)^{2\left(d/2 - u_k\right)}}\dfrac{1}{x_k^{2u_k}}\times
    \dfrac{1}{\left(x_L - y\right)^{2\left(d/2 - u_{L}\right)}}\dfrac{1}{y^{2u_{L}}},
\end{equation*}
where $x_0=x$. This integral corresponds to the so-called Feynman ladder diagram. Formula~(\ref{L}) gives a closed form expression, containing only one integration over $\nu$ and summation over $n$.

Also, if we recall relation (\ref{geg}), we can rewrite formula (\ref{L}) using the Gegenbauer polynomials 
\begin{equation*}
Q_{u_1\ldots u_{L}}\left(x, y\right) = 
\Gamma\left(d/2 - 1\right)\sum_{n \ge 0}\left(d/2 - 1 + n\right)\int_{\mathbb{R}}d\nu\dfrac{C_n^{\left(d/2 - 1\right)}\left(\hat{x}\cdot\hat{y}\right)}{x^{2\left(d/4 + i\nu\right)}y^{2\left(d/4 - i\nu\right)}}
\prod\limits_{k = 1}^{L}\tau\left(u_k,\nu,n\right).
\end{equation*}

\end{document}